# Pure plate bending in couple stress theories

Ali R. Hadjesfandiari, Arezoo Hajesfandiari, Gary F. Dargush

*Department of Mechanical and Aerospace Engineering*
*University at Buffalo, State University of New York, Buffalo, NY 14260 USA*

ah@buffalo.edu, ah62@buffalo.edu, gdargush@buffalo.edu

June 1, 2016

**Abstract**

In this paper, we examine the pure bending of plates within the framework of modified couple stress theory (M-CST) and consistent couple stress theory (C-CST). In this development, it is demonstrated that M-CST does not describe pure bending of a plate properly. Particularly, M-CST predicts no couple-stresses and no size effect for the pure bending of the plate into a spherical shell. This contradicts our expectation that couple stress theory should predict some size effect for such a deformation pattern. Therefore, this result clearly demonstrates another inconsistency of indeterminate symmetric modified couple stress theory (M-CST), which is based on considering the symmetric torsion tensor as the curvature tensor. On the other hand, the fully determinate skew-symmetric consistent couple stress theory (C-CST) predicts results for pure plate bending that tend to agree with mechanics intuition and experimental evidence. Particularly, C-CST predicts couple-stresses and size effects for the pure bending of the plate into a spherical shell, which represents an additional illustration of its consistency.

## 1. Introduction

Although the indeterminate symmetric modified couple stress theory (M-CST) (Yang et al., 2002) has been extensively used to analyze bending of structural elements, such as beams, plates and shells, its validity for the proper representation of bending deformation has not been examined carefully. Since this theory is based on considering the symmetric torsion tensor as the curvature tensor, the resulting symmetric couple-stresses create torsion or anticlastic deformation with negative Gaussian curvature for surface elements of a continuum. As a result, this theory cannot describe the bending of these structural elements properly. Particularly, this theory predicts no couple-stresses and no size effects for the pure bending deformation of a plate into a spherical shell. This result, which contradicts our general mechanics based understanding and small-scale bending experiments, demonstrates another inconsistency of the modified couple stress theory (M-CST).

On the other hand, the consistent couple stress theory (C-CST) (Hadjesfandiari and Dargush, 2011) is based on the skew-symmetric couple-stress and mean curvature tensors. As a result, the skew-symmetric couple-stresses create ellipsoidal cap-like deformation with positive Gaussian curvature for surface elements of a continuum. Therefore, this theory can describe the bending of structural elements, such as plates and shells properly. Specifically, this theory predicts some couple-stresses and significant size effect for the pure bending deformation of a plate into a spherical shell. This result agrees with our expectations, and clearly demonstrates another facet of the inherent consistency of this couple stress theory (C-CST).

In this paper, we show that the formulation of pure bending of a plate plays an important role in investigating the validity of M-CST versus C-CST. Although elements of M-CST and C-CST are based on the work of Mindlin and Tiersten (1962) and Koiter (1964), they cannot be taken as special cases of the original indeterminate Mindlin-Tiersten-Koiter couple stress theory (MTK-CST). This is because the curvature tensors are different from scratch in these theories. Although we do not present the pure bending of plates in MTK-CST in detail, the results for infinitesimal linear elasticity can be obtained by linear combination of the results from M-CST and C-CST. For further discussion on different aspects of couple stress theories, the reader is referred to Hadjesfandiari and Dargush (2011, 2015a,b, 2016).

The remainder of the paper is organized as follows. In Section 2, we consider the kinematics of pure bending deformation of plates. Next, in Section 3, we give a brief review of the general couple stress theory in small deformation solid mechanics. Section 4 considers pure plate bending in isotropic classical (Cauchy) elasticity, modified couple stress isotropic elasticity, and consistent couple stress isotropic elasticity in separate subsections. Then, Section 5 presents three special cases of plate bending deformations: spherical, cylindrical and equal curvature anticlastic deformations within the framework of the modified and consistent couple stress theories. These investigations clearly demonstrate the inconsistency of the modified couple stress theory (M-CST). Section 6 contains a brief discussion and an overall comparison of the presented theories, and demonstrates the inconsistency of some general plate bending M-CST based formulations. Finally, Section 7 provides several overall conclusions.

## 2. Kinematics of pure plate bending

Consider a flat plate of thickness $h$, with the middle plane $x_1 x_2$ before deformation, as shown in Fig. 1.

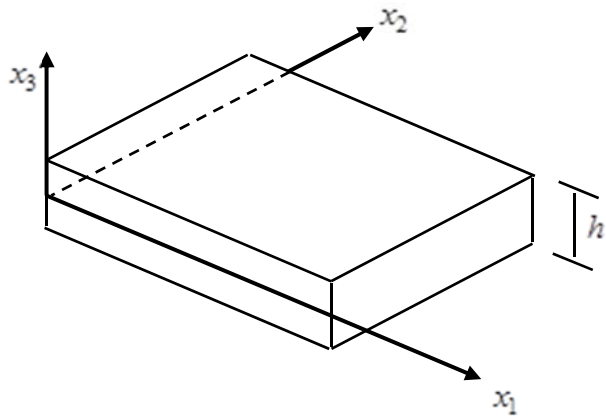

**Fig. 1.** Geometry and coordinate system of a plate.

Under some external loading, the plate undergoes deformation specified by the displacement field $u_i$. In this paper, we investigate the pure bending deformation of this plate into a shell within different elasticity theories. Let $R_1$ and $R_2$ denote the constant radius of curvatures of the middle deflection surface of the shell in planes parallel to $x_1 x_3$ and $x_2 x_3$, respectively, as shown in Fig. 2.

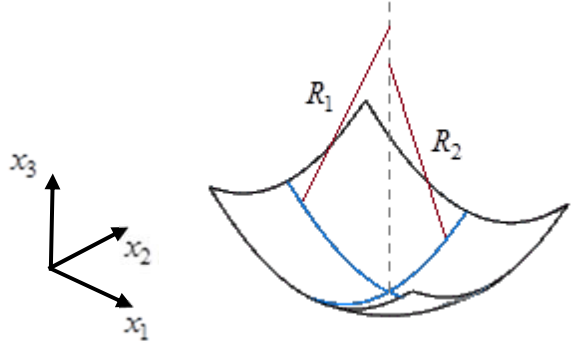

**Fig. 2.** Geometry of deformed middle plane of the plate in pure bending.

For positive curvatures in the $x_3$ direction, $R_1$ and $R_2$ are positive. In general, the middle surface is ellipsoidal, cylindrical or hyperboloidal depending on the values of $R_1$ and $R_2$. It turns out that the concept of Gaussian curvature of the middle surface of plate, defined as

$$K = \frac{1}{R_1 R_2} \tag{1}$$

is very suitable for this classification. We notice

$$K > 0 \qquad w \text{ is ellipsoidal} \tag{2a}$$

$$K = 0 \qquad w \text{ is cylindrical} \tag{2b}$$

$$K < 0 \qquad w \text{ is hyperboloidal (saddle surface)} \tag{2c}$$

where $w$ is the deflection of the middle surface in the $x_3$ direction. In small deformation theory, the middle deflected surface can be approximated as

$$w = u_3(x_1, x_2, x_3 = 0) = \frac{1}{2R_1} x_1^2 + \frac{1}{2R_2} x_2^2 \tag{3}$$

Based on the values of $R_1$ and $R_2$, this approximates the middle deflection surface as

$$K > 0 \quad \text{elliptic paraboloid} \tag{4a}$$

$$K = 0 \quad \text{parabolic cylinder} \tag{4b}$$

$$K < 0 \quad \text{hyperbolic paraboloid} \tag{4c}$$

## 3. Couple stress theory

In couple stress theory, the interaction in the body is represented by force-stress $\sigma_{ij}$ and couple-stress $\mu_{ij}$ tensors. The force and moment balance equations for general couple stress theory under quasi-static conditions in the absence of body forces are written, respectively, as:

$$\sigma_{ji,j} = 0 \tag{5}$$

$$\mu_{ji,j} + \varepsilon_{ijk}\sigma_{jk} = 0 \tag{6}$$

where $\varepsilon_{ijk}$ is the Levi-Civita alternating symbol and indices after commas represent derivatives with respect spatial coordinates.

In infinitesimal deformation theory, the displacement vector field $u_i$ is so small that the infinitesimal strain and rotation tensors are defined as

$$e_{ij} = u_{(i,j)} = \frac{1}{2}\left(u_{i,j} + u_{j,i}\right) \tag{7}$$

$$\omega_{ij} = u_{[i,j]} = \frac{1}{2}\left(u_{i,j} - u_{j,i}\right) \tag{8}$$

respectively. Since the tensor $\omega_{ij}$ is skew-symmetrical, one can introduce its corresponding dual rotation vector as

$$\omega_i = \frac{1}{2}\varepsilon_{ijk}\omega_{kj} = \frac{1}{2}\varepsilon_{ijk}u_{k,j} \tag{9}$$

where $\varepsilon_{ijk}$ is the Levi-Civita alternating symbol.

The infinitesimal torsion and mean curvature tensors are also defined as

$$\chi_{ij} = \omega_{(i,j)} = \frac{1}{2}\left(\omega_{i,j} + \omega_{j,i}\right) \tag{10}$$

$$\kappa_{ij} = \omega_{[i,j]} = \frac{1}{2}\left(\omega_{i,j} - \omega_{j,i}\right) \tag{11}$$

Since the mean curvature tensor $\kappa_{ij}$ is also skew-symmetrical, we can define its corresponding dual mean curvature vector as

$$\kappa_i = \frac{1}{2}\varepsilon_{ijk}\kappa_{kj} \tag{12}$$

This can also be expressed as

$$\kappa_i = \frac{1}{2}\omega_{ji,j} = \frac{1}{4}\left(u_{j,ji} - \nabla^2 u_i\right) \tag{13}$$

It should be noticed that the symmetric torsion tensor $\chi_{ij}$ represents the pure twist of the material (Hadjesfandiari and Dargush, 2011). In principal coordinates, this tensor becomes diagonal representing pure torsional deformations. On the other hand, the skew-symmetric mean curvature tensor $\kappa_{ij}$ represents the pure bending of the material.

In the following section, we examine the pure bending deformation of the plate in the linear isotropic classical and couple stress theories of elasticity. From physical experience, we expect that the size effects of pure bending of an isotropic elastic plate in planes parallel to $x_1 x_3$ and $x_2 x_3$ should add or cancel each other if the Gaussian curvature is positive or negative, respectively. Therefore, we anticipate some size effects for pure bending of the plate into a spherical shell with $R_1 = R_2$, but no size effects for bending into a pure equal curvature anticlastic saddle shell having $R_1 = -R_2$. However, the results from modified couple stress theory (M-CST) and the consistent couple stress theory (C-CST) contradict each other for these deformations, as will be shown.

## 4. Pure isotropic elastic plate bending

### *4.1. Classical (Cauchy) elasticity theory*

In classical or Cauchy elasticity, there are no couple-stresses and the force-stresses are symmetric, that is

$$\mu_{ij} = 0 \quad , \quad \sigma_{ji} = \sigma_{ij} \tag{14a,b}$$

The constitutive relation for a linear isotropic elastic material is

$$\sigma_{ij} = \sigma_{(ij)} = \lambda e_{kk} \delta_{ij} + 2G e_{ij} \tag{15}$$

where the moduli $\lambda$ and $G$ are the Lamé constants for isotropic media in Cauchy elasticity, and $G$ is also referred to as the shear modulus. These two constants are related by

$$\lambda = 2G \frac{\nu}{1-2\nu} \tag{16}$$

where $\nu$ is Poisson's ratio. In addition, we have the relations

$$\lambda = \frac{\nu E}{(1+\nu)(1-2\nu)}, \quad G = \frac{E}{2(1+\nu)}, \quad 3\lambda + 2G = \frac{E}{1-2\nu} \tag{17a-c}$$

where $E$ is Young's modulus of elasticity.

For the pure bending of a plate in classical isotropic elasticity, the displacement components up to an arbitrary rigid body motion are

$$u_1 = -\frac{1}{R_1} x_1 x_3 \tag{18a}$$

$$u_2 = -\frac{1}{R_2} x_2 x_3 \tag{18b}$$

$$u_3 = \frac{1}{2R_1} x_1^2 + \frac{1}{2R_2} x_2^2 + \frac{\nu}{2(1-\nu)} \left( \frac{1}{R_1} + \frac{1}{R_2} \right) x_3^2 \tag{18c}$$

We notice that there are no shear strain components for pure bending deformation under consideration here, that is

$$e_{12} = e_{13} = e_{23} = 0 \tag{19}$$

The non-zero components of the strain tensor are the normal strain components

$$e_{11} = -\frac{1}{R_1} x_3 \tag{20a}$$

$$e_{22} = -\frac{1}{R_2} x_3 \tag{20b}$$

$$e_{33} = \frac{\nu}{1-\nu}\left(\frac{1}{R_1} + \frac{1}{R_2}\right) x_3 \tag{20c}$$

Therefore, the non-zero force-stress components are written as

$$\sigma_{11} = -\frac{2\mu}{1-\nu}\left(\frac{1}{R_1} + \frac{\nu}{R_2}\right) x_3 = -\frac{E}{1-\nu^2}\left(\frac{1}{R_1} + \frac{\nu}{R_2}\right) x_3 \tag{21a}$$

$$\sigma_{22} = -\frac{2\mu}{1-\nu}\left(\frac{\nu}{R_1} + \frac{1}{R_2}\right) x_3 = -\frac{E}{1-\nu^2}\left(\frac{1}{R_1} + \frac{\nu}{R_2}\right) x_3 \tag{21b}$$

For the force-stress bending moments $M_1$ and $M_2$ per unit length on the edges of the plate parallel to the $x_2$ and $x_1$ axes, we obtain

$$M_1 = -\int_{-h/2}^{h/2} \sigma_{11}(z) z\, dz = D\left(\frac{1}{R_1} + \frac{\nu}{R_2}\right) \tag{22a}$$

$$M_2 = -\int_{-h/2}^{h/2} \sigma_{22}(z) z\, dz = D\left(\frac{\nu}{R_1} + \frac{1}{R_2}\right) \tag{22b}$$

where $D$ is the classical flexural rigidity of the plate defined as

$$D = \frac{1}{6}\frac{Gh^3}{1-v} = \frac{1}{12}\frac{Eh^3}{1-v^2} \qquad (23)$$

We have used the sign convention such that the bending moments $M_1$ and $M_2$ per unit length are positive if they create positive curvature $R_1$ and $R_2$ for the middle plane in the $x_3$ direction, as shown in Fig.3.

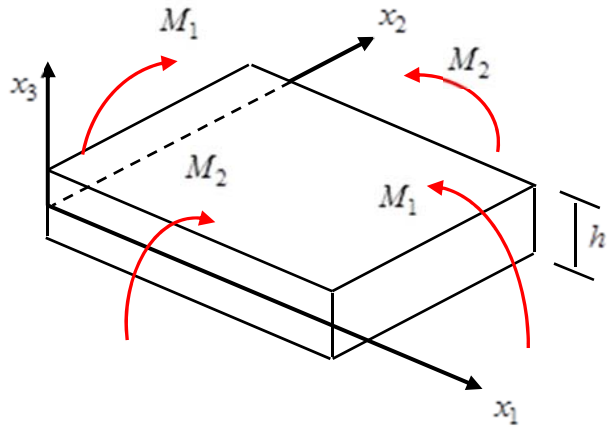

**Fig. 3.** Positive bending moments in plate theory.

In the following subsections, we investigate the pure bending deformation of this plate in different couple stress theories. These are isotropic elasticity based on the indeterminate symmetric modified couple stress theory (M-CST) and the determinate skew-symmetric consistent couple stress theory (C-CST). It should be noticed that the classical solutions Eq. (18) for displacement components still satisfy the governing equations in these theories. Therefore, in the next two subsections, we assume the same pure bending displacement components from Eq. (18) in these couple stress theories and investigate the corresponding stress distributions. For this we need to derive the expressions for the rotation vector, torsion tensor and mean curvature tensor as follows.

The components of the rotation vector are

$$\omega_1 = \omega_{32} = \frac{1}{R_2} x_2 \tag{24a}$$

$$\omega_2 = \omega_{13} = -\frac{1}{R_1} x_1 \tag{24b}$$

$$\omega_3 = \omega_{21} = 0 \tag{24c}$$

For the torsion tensor, we obtain

$$\left[\chi_{ij}\right] = \begin{bmatrix} 0 & \chi_{12} & 0 \\ \chi_{21} & 0 & 0 \\ 0 & 0 & 0 \end{bmatrix} \tag{25}$$

where the non-zero components $\chi_{12}$ and $\chi_{21}$ are

$$\chi_{12} = \chi_{21} = -\frac{1}{2}\left(\frac{1}{R_1} - \frac{1}{R_2}\right) \tag{26}$$

On the other hand, the mean curvature tensor becomes

$$\left[\kappa_{ij}\right] = \begin{bmatrix} 0 & \kappa_{12} & 0 \\ \kappa_{21} & 0 & 0 \\ 0 & 0 & 0 \end{bmatrix} \tag{27}$$

where the non-zero components $\kappa_{12}$ and $\kappa_{21}$ are

$$\kappa_{12} = \frac{1}{2}\left(\frac{1}{R_1} + \frac{1}{R_2}\right) \tag{28a}$$

$$\kappa_{21} = -\frac{1}{2}\left(\frac{1}{R_1} + \frac{1}{R_2}\right) \tag{28b}$$

We notice that the absolute values of $\kappa_{12}$ and $\kappa_{21}$ represent the average curvature of the middle deflection surface of the plate. This observation justifies use of the term "mean curvature tensor" for the skew-symmetric tensor $\kappa_{ij}$.

## 4.2. Symmetric modified couple stress theory (M-CST)

In the modified couple stress theory originally proposed by Yang et al. (2002), the couple-stress tensor is symmetric, that is

$$\mu_{ji} = \mu_{ij} \qquad (29)$$

and the curvature tensor is given by the symmetric tensor

$$\chi_{ij} = \chi_{ji} = \frac{1}{2}\left(\omega_{i,j} + \omega_{j,i}\right) \qquad (30)$$

We notice that in this theory, the normal couple-stress components on plane element surfaces create torsion, and tangential components deform these plane elements into anticlastic surfaces. For example, the tangential couple stresses $\mu_{12}$ and $\mu_{21}$ on the edges of plane surface elements parallel to $x_1 x_2$, as shown schematically in Fig. 4, create anticlastic deformation (a saddle surface) in the $x_3$ direction. However, this is inconsistent with our expectation that in a correct couple stress theory, the couple stresses $\mu_{12}$ and $\mu_{21}$ on the edges of plane surface elements should create a cap-like surface deformation with positive Gaussian curvature.

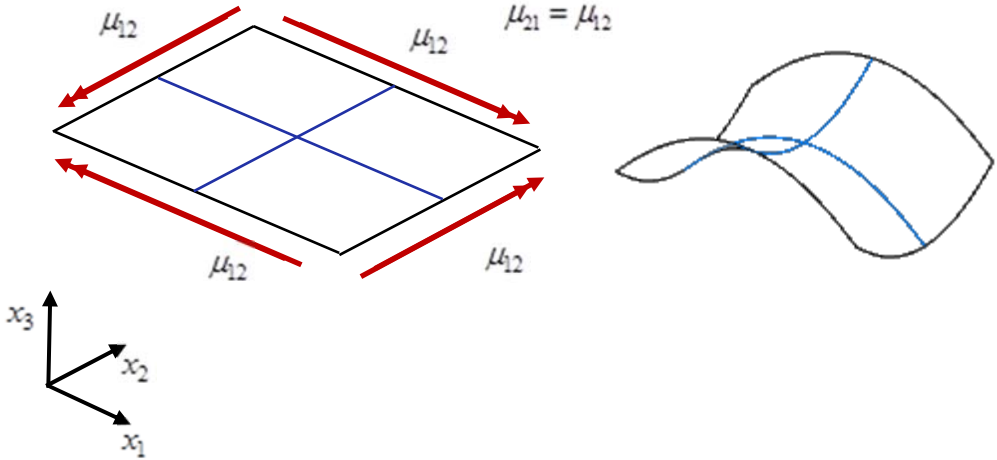

**Fig. 4.** Effect of couple-stress components $\mu_{12}$ and $\mu_{21}$ in M-CST: couple-stresses deform the surface plane elements parallel to $x_1 x_2$ plane to an anticlastic (saddle) surface element with negative Gaussian curvature.

We should mention that in M-CST the symmetric couple-stress tensor $\mu_{ij}$ creates torsional deformation along principal axes of this tensor. For clarification, let us define a new coordinate system $x'_1 x'_2 x'_3$ by rotating the coordinate system $x_1 x_2 x_3$ 45 degrees around the $x_3$ axis, as illustrated in schematic form in Fig. 5.

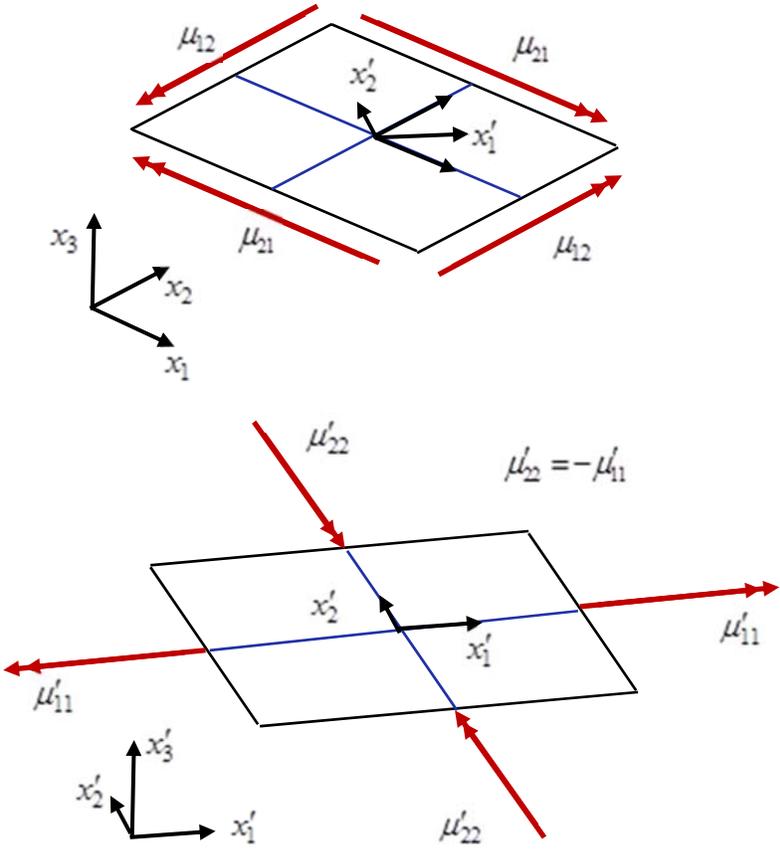

**Fig. 5.** In M-CST the bending components $\mu_{12}$ and $\mu_{21}$ in the original coordinate system $x_1 x_2 x_3$ transform into torsional components $\mu'_{11}$ and $\mu'_{22}$ in the new coordinate system $x'_1 x'_2 x'_3$.

The representation of the couple-stress tensors in this coordinate system is

$$\left[\mu'_{ij}\right] = \begin{bmatrix} \mu'_{11} & 0 & 0 \\ 0 & \mu'_{22} & 0 \\ 0 & 0 & 0 \end{bmatrix} \quad (31)$$

where

$$\mu'_{11} = \mu_{12}, \qquad \mu'_{22} = -\mu_{12} \quad (32a,b)$$

We notice that in the new coordinate system $x'_1 x'_2 x'_3$, the normal couple-stress components $\mu'_{11}$ and $\mu'_{22}$ create torsional deformation, not bending on plane element surfaces, as shown schematically in Fig. 5.

The constitutive relations for linear isotropic elastic materials are

$$\sigma_{(ij)} = \lambda e_{kk} \delta_{ij} + 2G e_{ij} \quad (33)$$

$$\mu_{ij} = Q \delta_{ij} + 8Gl^2 \chi_{ij} \quad (34)$$

where $\delta_{ij}$ is the Kronecker delta, and the constant $l$ is the characteristic material length in the modified couple stress theory (M-CST). We notice that the spherical part of the couple-stress tensor is indeterminate, where $Q$ is a pseudo-scalar.

It should be noticed that in presenting the modified couple stress theory (M-CST), we have used a different characteristic material length $l$ to have more similarity to the other couple stress theories. The relations in the modified couple stress theory (M-CST) in its original form (Yang et al., 2002) can be found by scaling

$$l \to l/2 \quad (35)$$

For the pure bending of a plate, the displacement, strain, rotation, force-stress tensors in M-CST are similar to those in classical elasticity given in Section 4.1.

In M-CST, the non-zero components of the symmetric tensor $\chi_{ij}$ are

$$\chi_{12} = \chi_{21} = -\frac{1}{2}\left(\frac{1}{R_1} - \frac{1}{R_2}\right) \tag{36}$$

As mentioned, the component $\chi_{12}$ represents the measure of deviation from sphericity of deforming planes parallel to $x_1 x_2$. This has dramatic consequences for M-CST theory.

The non-zero force and couple-stresses are written as

$$\sigma_{11} = -\frac{2\mu}{1-\nu}\left(\frac{1}{R_1} + \frac{\nu}{R_2}\right)x_3 = -\frac{E}{1-\nu^2}\left(\frac{1}{R_1} + \frac{\nu}{R_2}\right)x_3 \tag{37a}$$

$$\sigma_{22} = -\frac{2\mu}{1-\nu}\left(\frac{\nu}{R_1} + \frac{1}{R_2}\right)x_3 = -\frac{E}{1-\nu^2}\left(\frac{1}{R_1} + \frac{\nu}{R_2}\right)x_3 \tag{37b}$$

$$\mu_{12} = \mu_{21} = 8Gl^2 \chi_{21} = -4Gl^2\left(\frac{1}{R_1} - \frac{1}{R_2}\right) \tag{38}$$

Since there is no normal twist component on the boundaries of the plate (because $\chi_{11} = \chi_{22} = \chi_{33} = 0$), we have taken $Q = 0$.

We notice that the representations of the torsion and couple-stress tensors in the coordinate system $x'_1 x'_2 x'_3$ are

$$[\chi'_{ij}] = \begin{bmatrix} \chi'_{11} & 0 & 0 \\ 0 & \chi'_{22} & 0 \\ 0 & 0 & 0 \end{bmatrix}, \quad [\mu'_{ij}] = \begin{bmatrix} \mu'_{11} & 0 & 0 \\ 0 & \mu'_{22} & 0 \\ 0 & 0 & 0 \end{bmatrix} \tag{39a,b}$$

where

$$\chi'_{11} = \chi_{12} = -\frac{1}{2}\left(\frac{1}{R_1} - \frac{1}{R_2}\right), \quad \chi'_{22} = -\chi_{12} = \frac{1}{2}\left(\frac{1}{R_1} - \frac{1}{R_2}\right) \tag{40a,b}$$

$$\mu'_{11} = \mu_{12} = 4Gl^2\left(\frac{1}{R_1} - \frac{1}{R_2}\right), \quad \mu'_{22} = -\mu_{12} = -4Gl^2\left(\frac{1}{R_1} - \frac{1}{R_2}\right) \tag{41a,b}$$

These representations clearly show that modified couple stress theory (M-CST) is based on torsional deformation, rather than characterizing bending measures of deformation.

For the force-stress bending moments $M_{\sigma 1}$ and $M_{\sigma 2}$ per unit length on the edges of the plate parallel to the $x_2$ and $x_1$ axes, we still obtain the classical results

$$M_{\sigma 1} = -\int_{-h/2}^{h/2} \sigma_{11}(z) z\, dz = D\left(\frac{1}{R_1} + \frac{\nu}{R_2}\right) \tag{42a}$$

$$M_{\sigma 2} = -\int_{-h/2}^{h/2} \sigma_{22}(z) z\, dz = D\left(\frac{\nu}{R_1} + \frac{1}{R_2}\right) \tag{42b}$$

For the couple-stress bending moments $M_{\mu 1}$ and $M_{\mu 2}$ per unit length on the edges of the plate parallel to the $x_2$ and $x_1$ axes, we obtain

$$M_{\mu 1} = -\int_{-h/2}^{h/2} \mu_{12}(z)\, dz = -\mu_{12} h = 4Gl^2\left(\frac{1}{R_1} - \frac{1}{R_2}\right) \tag{43a}$$

$$M_{\mu 2} = \int_{-h/2}^{h/2} \mu_{21}(z)\, dz = \mu_{21} h = -4Gl^2 h\left(\frac{1}{R_1} - \frac{1}{R_2}\right) \tag{43b}$$

Therefore, the total bending moments $M_1$ and $M_2$ per unit length on these same edges of the plate parallel to the $x_2$ and $x_1$ axes, we obtain

$$\begin{aligned} M_1 = M_{\sigma 1} + M_{\mu 1} &= D\left(\frac{1}{R_1} + \frac{\nu}{R_2}\right) + 4Gl^2 h\left(\frac{1}{R_1} - \frac{1}{R_2}\right) \\ &= D\left\{\left(\frac{1}{R_1} + \frac{\nu}{R_2}\right) + 24(1-\nu)\frac{l^2}{h^2}\left(\frac{1}{R_1} - \frac{1}{R_2}\right)\right\} \end{aligned} \tag{44a}$$

$$M_2 = M_{\sigma 2} + M_{\mu 2} = D\left(\frac{\nu}{R_1} + \frac{1}{R_2}\right) + 4Gl^2 h\left(\frac{1}{R_2} - \frac{1}{R_1}\right)$$
$$= D\left\{\left(\frac{\nu}{R_1} + \frac{1}{R_2}\right) + 24(1-\nu)\frac{l^2}{h^2}\left(\frac{1}{R_2} - \frac{1}{R_1}\right)\right\} \quad (44b)$$

We notice the opposite signs in Eq. (43) for the coefficients of the terms containing $R_1$ and $R_2$ in couple-stress bending moments $M_{\mu 1}$ and $M_{\mu 2}$. This is in contrast with the same signs for the coefficients of terms containing $R_1$ and $R_2$ in Eq. (42) for force-stress bending moments $M_{\sigma 1}$ and $M_{\sigma 2}$. This means that the force-stress bending moments $M_{\sigma 1}$ and $M_{\sigma 2}$ bend the plate in the $x_3$ direction in the same sense, but the couple-stress bending moments $M_{\mu 1}$ and $M_{\mu 2}$ create anticlastic deformation. This result does not agree with our intuition and is not consistent with physical reality. For a pure bending, we expect $R_1$ and $R_2$ to have the same signs in expressions for $M_\sigma$ and $M_\mu$. This clearly demonstrates an inconsistency of M-CST.

### 4.3. Skew-symmetric consistent couple stress theory (C-CST)

In this theory (Hadjesfandiari and Dargush, 2011), the couple stress tensor is skew-symmetric, that is

$$\mu_{ji} = -\mu_{ij} \quad (45)$$

and the curvature tensor is the skew-symmetric mean curvature tensor

$$\kappa_{ij} = \frac{1}{2}\left(\omega_{i,j} - \omega_{j,i}\right) \quad (46)$$

We notice that in this theory the couple-stress components deform the plane element surfaces into ellipsoidal surfaces with positive Gaussian curvature. For example, the couple stresses $\mu_{12}$ and $\mu_{21}$ on the edges of plane surface elements parallel to $x_1 x_2$, create an ellipsoidal surface in the negative $x_3$ direction with positive Gaussian curvature. This is what we intuitively expect from a consistent couple stress theory, which is shown schematically in Fig. 6

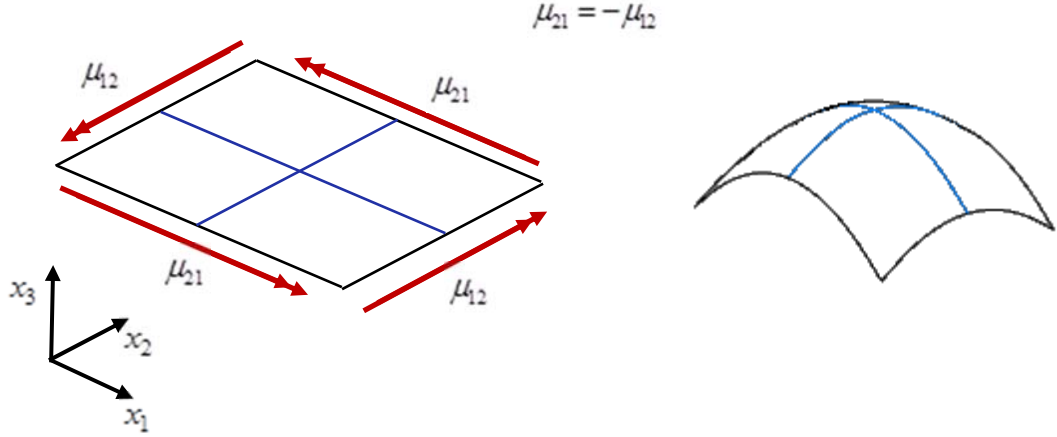

**Fig. 6.** Effect of couple-stress components $\mu_{12}$ and $\mu_{21}$ in C-CST: couple-stresses deform the surface plane elements parallel to $x_1 x_2$ plane to an ellipsoid cap-like surface element with positive Gaussian curvature.

The constitutive relations for a linear isotropic elastic material within C-CST are

$$\sigma_{(ij)} = \lambda e_{kk} \delta_{ij} + 2G e_{ij} \tag{47}$$

$$\mu_{ij} = -8Gl^2 \kappa_{ij} \tag{48}$$

where there is no indeterminacy. The parameter $l$ is the characteristic material length in the consistent couple stress theory (C-CST).

For the pure plate bending in this theory, the displacement, strain, rotation, and force-stress tensors are identical to those in classical elasticity, obtained in Section 4.1.

In C-CST, the non-zero components of the skew-symmetric mean curvature tensor $\kappa_{ij}$ are

$$\kappa_{12} = -\kappa_3 = \frac{1}{2}\left(\frac{1}{R_1} + \frac{1}{R_2}\right) \tag{49a}$$

$$\kappa_{21} = \kappa_3 = -\frac{1}{2}\left(\frac{1}{R_1} + \frac{1}{R_2}\right) \tag{49b}$$

which represent the mean curvature of the middle deflection surface of the plate.

Therefore, the non-zero force and couple-stresses are written as

$$\sigma_{11} = -\frac{2G}{1-v}\left(\frac{1}{R_1} + \frac{v}{R_2}\right)x_3 = -\frac{E}{1-v^2}\left(\frac{1}{R_1} + \frac{v}{R_2}\right)x_3 \tag{50a}$$

$$\sigma_{22} = -\frac{2G}{1-v}\left(\frac{v}{R_1} + \frac{1}{R_2}\right)x_3 = -\frac{E}{1-v^2}\left(\frac{1}{R_1} + \frac{v}{R_2}\right)x_3 \tag{50b}$$

$$\mu_{12} = -8Gl^2\kappa_{12} = -4Gl^2\left(\frac{1}{R_1} + \frac{1}{R_2}\right) \tag{51a}$$

$$\mu_{21} = -\mu_{21} = -8Gl^2\kappa_{21} = 4Gl^2\left(\frac{1}{R_1} + \frac{1}{R_2}\right) \tag{51b}$$

For the force-stress bending moments $M_{\sigma 1}$ and $M_{\sigma 2}$, we still have the classical results

$$M_{\sigma 1} = -\int_{-h/2}^{h/2}\sigma_{11}(z)zdz = D\left(\frac{1}{R_1} + \frac{v}{R_2}\right) \tag{52a}$$

$$M_{\sigma 2} = -\int_{-h/2}^{h/2}\sigma_{22}(z)zdz = D\left(\frac{v}{R_1} + \frac{1}{R_2}\right) \tag{52b}$$

Meanwhile, for the couple-stress bending moments $M_{\mu 1}$ and $M_{\mu 2}$, we obtain

$$M_{\mu 1} = -\int_{-h/2}^{h/2}\mu_{12}(z)dz = -\mu_{12}h = 4Gl^2h\left(\frac{1}{R_1} + \frac{1}{R_2}\right) \tag{53a}$$

$$M_{\mu 2} = \int_{-h/2}^{h/2}\mu_{21}(z)dz = \mu_{21}h = 4Gl^2h\left(\frac{1}{R_1} + \frac{1}{R_2}\right) \tag{53b}$$

Therefore, for the total bending moments $M_1$ and $M_2$, we obtain

$$M_1 = M_{\sigma 1} + M_{\mu 1} = D\left(\frac{1}{R_1} + \frac{v}{R_2}\right) + 4Gl^2h\left(\frac{1}{R_1} + \frac{1}{R_2}\right)$$
$$= D\left\{\left(\frac{1}{R_1} + \frac{v}{R_2}\right) + 24(1-v)\frac{l^2}{h^2}\left(\frac{1}{R_1} + \frac{1}{R_2}\right)\right\} \quad (54a)$$

$$M_2 = M_{\sigma 2} + M_{\mu 2} = D\left(\frac{v}{R_1} + \frac{1}{R_2}\right) + 4Gl^2h\left(\frac{1}{R_1} + \frac{1}{R_2}\right)$$
$$= D\left\{\left(\frac{v}{R_1} + \frac{1}{R_2}\right) + 24(1-v)\frac{l^2}{h^2}\left(\frac{1}{R_1} + \frac{1}{R_2}\right)\right\} \quad (54b)$$

We notice the same signs for coefficients of terms containing $R_1$ and $R_2$ in Eqs. (52) and (53) for force- and couple-stress bending moments $M_\sigma$ and $M_\mu$. This means that the bending moments $M_\sigma$ and $M_\mu$ in orthogonal directions $x_1$ and $x_2$ bend the plate in the $x_3$ direction in the same sense. This agrees with our expectation from physical reality, and shows the consistency of C-CST.

## 5. Special cases of pure plate bending

In this section, we consider special cases of plate bending deformations: spherical, cylindrical and equal curvature anticlastic deformations within the framework of modified and consistent couple stress theories. While this development reveals more clearly the inconsistency of the modified couple stress theory (M-CST), it also demonstrates the consistency of C-CST.

### 5.1. Spherical bending of the flat plate $R_1 = R_2$

For this first special case, the plate is bent to a spherical shell, as shown in Figure 7.

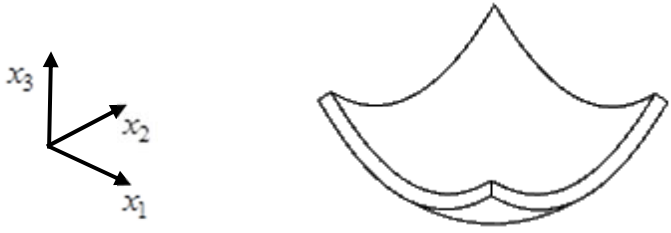

**Fig. 7.** Bending of plate to a pure spherical shell.

In small deformation theory, the displacement field for this deformation becomes

$$u_1 = -\frac{1}{R_1} x_1 x_3 \tag{55a}$$

$$u_2 = -\frac{1}{R_1} x_2 x_3 \tag{55b}$$

$$u_3 = \frac{1}{2R_1}\left(x_1^2 + x_2^2\right) + \frac{\nu}{1-\nu}\frac{1}{R_1} x_3^2 \tag{55c}$$

Therefore, the middle plane of the plate is bent to a spherical surface, which has been approximated by the paraboloid

$$w = u_3(x_1, x_2, x_3 = 0) = \frac{1}{2R_1}\left(x_1^2 + x_2^2\right) \tag{56}$$

In both couple stress theories, the non-zero force-stress components become

$$\sigma_{11} = \sigma_{22} = -2G\frac{1+\nu}{1-\nu}\frac{1}{R_1} x_3 \tag{57}$$

We notice that for this deformation the torsion tensor vanishes, because

$$\chi_{12} = \chi_{21} = 0 \tag{58}$$

However, for this deformation the non-zero components of the mean curvature tensor are

$$\kappa_{12} = \frac{1}{R_1} \qquad \kappa_{21} = -\frac{1}{R_1} \tag{59a,b}$$

These have dramatic consequences on the total bending moments $M_1$ and $M_2$ in couple stress theories under consideration as follows.

### 5.1.1. M-CST theory

In this theory, couple-stresses vanish, that is

$$\mu_{12} = \mu_{21} = 0 \tag{60}$$

and the solution reduces to the classical solution. For bending moments we have

$$M_{\sigma 1} = D(1+v)\frac{1}{R_1}, \qquad M_{\sigma 2} = D(1+v)\frac{1}{R_1} \tag{61a,b}$$

$$M_{\mu 1} = 0, \qquad M_{\mu 2} = 0 \tag{62a,b}$$

$$M_1 = M_{\sigma 1} + M_{\mu 1} = D(1+v)\frac{1}{R_1}, \qquad M_2 = M_{\sigma 2} + M_{\mu 2} = D(1+v)\frac{1}{R_1} \tag{63a,b}$$

Since there is no deviation from sphericity $\chi_{12} = 0$ for this deformation, there is no torsion tensor and no couple-stresses. Therefore, M-CST predicts no size effect for bending of the plate into a spherical shell, $R_1 = R_2$. Contrary to our expectation, the size effects from bending in directions $x_1$ and $x_2$ cancel each other. This result clearly demonstrates that the modified couple stress theory (M-CST) predicts inconsistent and non-physical results.

*5.1.2. C-CST theory*

Since there is non-zero mean curvatures for this deformation, the consistent couple stress theory predicts the couple-stresses

$$\mu_{12} = -8Gl^2 \frac{1}{R_1}, \quad \mu_{21} = 8Gl^2 \frac{1}{R_1} \tag{64a,b}$$

For bending moments, we have

$$M_{\sigma 1} = D(1+v)\frac{1}{R_1}, \quad M_{\sigma 2} = D(1+v)\frac{1}{R_1} \tag{65a,b}$$

$$M_{\mu 1} = 8Gl^2 h \frac{1}{R_1}, \quad M_{\mu 2} = 8Gl^2 h \frac{1}{R_1} \tag{66a,b}$$

$$M_1 = M_{\sigma 1} + M_{\mu 1} = D\left[1+v+48(1-v)\frac{l^2}{h^2}\right]\frac{1}{R_1} \tag{67a}$$

$$M_2 = M_{\sigma 2} + M_{\mu 2} = D\left[1+v+48(1-v)\frac{l^2}{h^2}\right]\frac{1}{R_1} \tag{67b}$$

We notice that the bending moments $M_\sigma$ and $M_\sigma$ in the orthogonal directions $x_1$ and $x_2$ bend the plate in the $x_3$ direction in the same sense. Therefore, the consistent couple stress theory (C-CST) predicts some size effect for bending of a plate into a spherical shell, $R_1 = R_2$. As we expected, the size effects from bending in directions $x_1$ and $x_2$ add together, which clearly demonstrates the consistency of C-CST.

*5.2. Cylindrical bending of the flat plate $R_2 = \infty$*

For the second special case, consider the plate bent into a circular cylindrical shell, as shown in Figure 8.

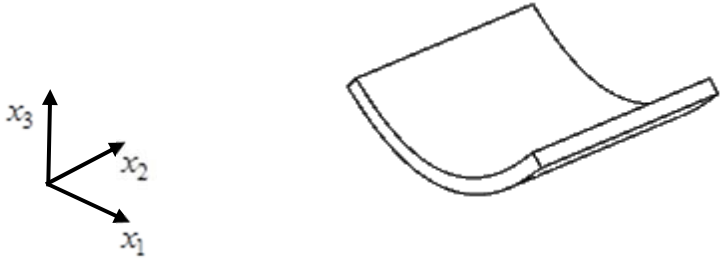

**Fig. 8.** Bending of plate to a cylindrical shell.

In small deformation theory, the displacement field for this deformation becomes

$$u_1 = -\frac{1}{R_1} x_1 x_3 \tag{68a}$$

$$u_2 = 0 \tag{68b}$$

$$u_3 = \frac{1}{2R_1} x_1^2 + \frac{\nu}{2(1-\nu)} \frac{1}{R_1} x_3^2 \tag{68c}$$

Therefore, the middle plane of the plate is bent to a circular cylindrical surface, which has been approximated by the parabolic cylinder

$$w = u_3(x_1, x_2, x_3 = 0) = \frac{1}{2R_1} x_1^2 \tag{69}$$

For this deformation, the non-zero force-stress components become

$$\sigma_{11} = -2G \frac{1}{1-\nu} \frac{1}{R} x_3, \quad \sigma_{22} = -2G \frac{\nu}{1-\nu} \frac{1}{R} x_3 \tag{70a,b}$$

We notice that for this deformation, there exist torsion and mean curvature components

$$\chi_{12} = \chi_{21} = -\frac{1}{2R_1} \tag{71}$$

$$\kappa_{12} = \frac{1}{2R_1}, \qquad \kappa_{21} = -\frac{1}{2R_1} \tag{72a,b}$$

Now we investigate the bending moments $M_1$ and $M_2$ in couple stress theories under consideration as follows.

### 5.2.1. M-CST theory

In this theory, the non-zero couple-stresses become

$$\mu_{12} = \mu_{21} = 8Gl^2 \chi_{21} = -4Gl^2 \frac{1}{R_1} \tag{73}$$

For bending moments, we have

$$M_{\sigma 1} = D\frac{1}{R_1}, \qquad M_{\sigma 2} = \nu D\frac{1}{R_1} \tag{74a,b}$$

$$M_{\mu 1} = 4Gl^2 h \frac{1}{R_1}, \qquad M_{\mu 2} = -4Gl^2 h \frac{1}{R_1} \tag{75a,b}$$

$$M_1 = M_{\sigma 1} + M_{\mu 1} = D\left[1 + 24(1-\nu)\frac{l^2}{h^2}\right]\frac{1}{R_1} \tag{76a}$$

$$M_2 = M_{\sigma 2} + M_{\mu 2} = D\left[\nu - 24(1-\nu)\frac{l^2}{h^2}\right]\frac{1}{R_1} \tag{76b}$$

We notice that while $M_{\sigma 1}$ and $M_{\sigma 2}$ bend the plate in the $x_3$ direction in the same sense, $M_{\mu 1}$ and $M_{\mu 2}$ create anticlastic deformation with negative Gaussian curvature. We can demonstrate the inconsistency of M-CST more clearly by noticing when the thickness of the plate decreases, such that

$$vD = 4Gl^2h \quad \text{or} \quad v - 24(1-v)\frac{l^2}{h^2} = 0 \qquad (77)$$

the total bending moment $M_2$ disappears, that is $M_2 = 0$. This is because $M_{\sigma 2}$ and $M_{\mu 2}$ cancel each other. Therefore, the critical thickness of the plate is

$$h = 2l\sqrt{\frac{6(1-v)}{v}} \qquad (78)$$

By decreasing the thickness $h$ further, the bending moment $M_2$ becomes negative for $R_1 > 0$. Therefore, we have

$$M_2 > 0 \quad \text{for} \quad h > 2l\sqrt{\frac{6(1-v)}{v}} \qquad (79a)$$

$$M_2 = 0 \quad \text{for} \quad h = 2l\sqrt{\frac{6(1-v)}{v}} \qquad (79b)$$

$$M_2 < 0 \quad \text{for} \quad h < 2l\sqrt{\frac{6(1-v)}{v}} \qquad (79c)$$

These non-physical results, which are displayed in Figure 9, are the result of an inconsistency of M-CST.

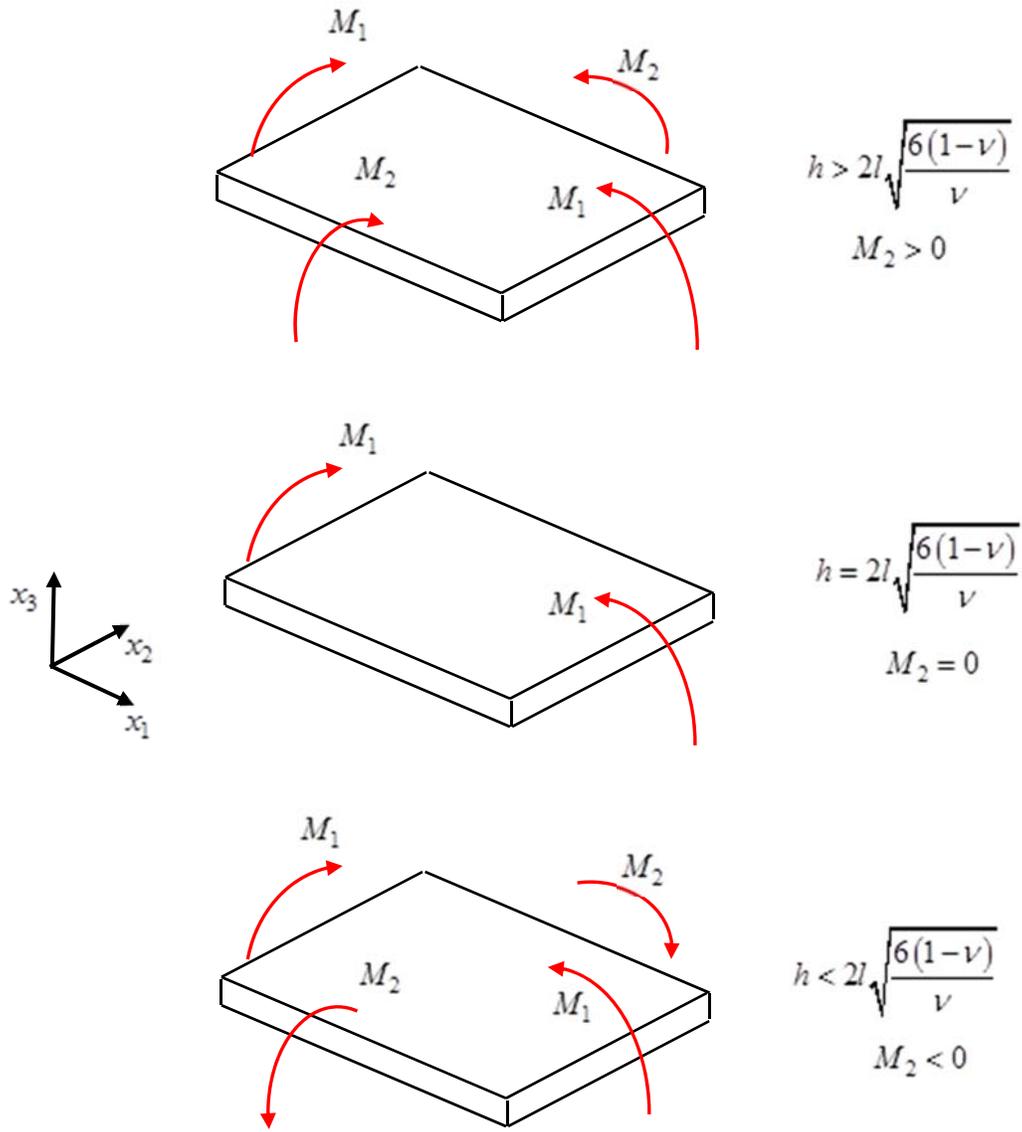

**Fig. 9.** Bending moments for cylindrical bending of the flat plate in M-CST for different thickness $h$.

*5.2.2. C-CST theory*

This theory predicts the following non-zero couple stresses:

$$\mu_{12} = -8Gl^2 \kappa_{12} = -4Gl^2 \frac{1}{R_1} \tag{80a}$$

$$\mu_{21} = -8Gl^2 \kappa_{21} = 4Gl^2 \frac{1}{R_1} \tag{80b}$$

while for bending moments, we have

$$M_{\sigma 1} = D\frac{1}{R_1}, \qquad M_{\sigma 2} = \nu D \frac{1}{R_1} \tag{81a,b}$$

$$M_{\mu 1} = 4Gl^2 h \frac{1}{R_1}, \qquad M_{\mu 2} = 4Gl^2 h \frac{1}{R_1} \tag{82a,b}$$

$$M_1 = M_{\sigma 1} + M_{\mu 1} = D\left[1 + 24(1-\nu)\frac{l^2}{h^2}\right]\frac{1}{R_1} \tag{83a}$$

$$M_2 = M_{\sigma 2} + M_{\mu 2} = D\left[\nu + 24(1-\nu)\frac{l^2}{h^2}\right]\frac{1}{R_1} \tag{83b}$$

We notice that the bending moments $M_\sigma$ and $M_\mu$ in the orthogonal directions $x_1$ and $x_2$ bend the plate in the $x_3$ direction in the same sense to create positive Gaussian curvature. This result again demonstrates the consistency of C-CST.

*5.3. Equal curvature anticlastic bending of the flat plate* $R_1 = -R_2$

In this final special case, the plate is bent to an equal curvature anticlastic shell, as shown in Figure 10.

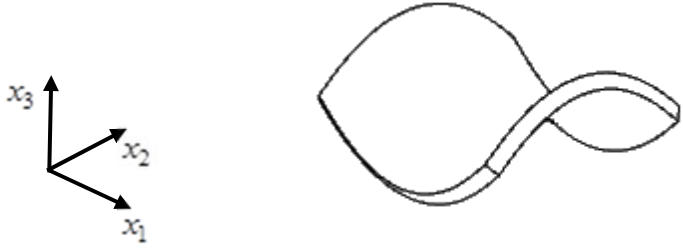

**Fig. 10.** Bending of plate to an equal curvature saddle shell.

The displacement field for this deformation becomes

$$u_1 = -\frac{1}{R_1} x_1 x_3 \tag{84a}$$

$$u_2 = \frac{1}{R_1} x_2 x_3 \tag{84b}$$

$$u_3 = \frac{1}{2R_1}\left(x_1^2 - x_2^2\right) \tag{84c}$$

Therefore, the middle plane of the plate is bent to an equal curvature anticlastic surface, which has been approximated by the hyperbolic paraboloid (saddle surface)

$$w = u_3\left(x_1, x_2, x_3 = 0\right) = \frac{1}{2R_1}\left(x_1^2 - x_2^2\right) \tag{85}$$

For this deformation, the non-zero force-stress components are

$$\sigma_{11} = -2G\frac{1}{R_1} x_3, \qquad \sigma_{22} = 2G\frac{1}{R_1} x_3 \tag{86a}$$

We notice that for this deformation the non-zero components of the torsion tensor are

$$\chi_{12} = \chi_{21} = -\frac{1}{R_1} \tag{87}$$

On the other hand, the mean curvature tensor vanishes, because

$$\kappa_{12} = 0, \quad \kappa_{21} = 0 \tag{88a,b}$$

We notice that this deformation is equivalent to a combination of two torsional deformations along 45 degree orthogonal directions from the coordinate axes $x_1$ and $x_2$ in the plane $x_1 x_2$, as explained in Section 4.2. Now we investigate the consequence of these on the bending moments $M_1$ and $M_2$ in the two couple stress theories under consideration.

### 5.3.1. M-CST theory

In this theory, the non-zero couple-stresses become

$$\mu_{12} = \mu_{21} = 8Gl^2 \chi_{21} = -8Gl^2 \frac{1}{R_1} \tag{89}$$

For the bending moments, we have

$$M_{\sigma 1} = D(1-v)\frac{1}{R_1}, \qquad M_{\sigma 2} = -D(1-v)\frac{1}{R_1} \tag{90a,b}$$

$$M_{\mu 1} = 8Gl^2 h \frac{1}{R_1}, \qquad M_{\mu 2} = -8Gl^2 h \frac{1}{R_1} \tag{91a,b}$$

$$M_1 = M_{\sigma 1} + M_{\mu 1} = D(1-v)\left(1 + 48\frac{l^2}{h^2}\right)\frac{1}{R_1} \tag{92a}$$

$$M_2 = M_{\sigma 2} + M_{\mu 2} = -D(1-v)\left(1 + 48\frac{l^2}{h^2}\right)\frac{1}{R_1} \tag{92b}$$

Contrary to our expectation, the size effects from bending in directions $x_1$ and $x_2$ add together. This result once again clearly demonstrates the inconsistency of M-CST.

*5.3.2. C-CST theory*

Since there is no mean curvatures for this deformation, this theory predicts no couple-stresses, that is

$$\mu_{12} = -8Gl^2\kappa_{12} = 0 \qquad \mu_{21} = -8Gl^2\kappa_{21} = 0 \tag{93a,b}$$

For the bending moments, we have

$$M_{\sigma 1} = D(1-\nu)\frac{1}{R_1} \qquad M_{\sigma 2} = -D(1-\nu)\frac{1}{R_1} \tag{94a,b}$$

$$M_{\mu 1} = 0 \qquad M_{\mu 2} = 0 \tag{95a,b}$$

$$M_1 = M_{\sigma 1} + M_{\mu 1} = D(1-\nu)\frac{1}{R_1} \tag{96a}$$

$$M_2 = M_{\sigma 2} + M_{\mu 2} = -D(1-\nu)\frac{1}{R_1} \tag{96b}$$

We notice that the solution has reduced to the classical solution. Therefore, consistent couple stress theory (C-CST) predicts no size effect for an equal curvature anticlastic deformation $R_2 = -R_1$. This agrees with our expectation that the size effects from bending in orthogonal directions $x_1$ and $x_2$ cancel each other, once again confirming the validity of C-CST in representing size-dependent mechanics.

## 6. Discussion

*6.1. M-CST versus C-CST in pure plate bending*

From the analysis of pure plate bending problems in the modified couple stress theory (M-CST) and the consistent couple stress theory (C-CST), we have obtained some contradictory results as follows.

Within modified couple stress theory (M-CST), when the pure bending deformation deviates from a pure spherical bending, where $\frac{1}{R_1} - \frac{1}{R_2}$ is non-zero, the couple-stress components appear and the flexural rigidity of the plate changes such that

$$M_1 = M_{\sigma 1} + M_{\mu 1} = D\left(\frac{1}{R_1} + \frac{\nu}{R_2}\right) + 4Gl^2 h\left(\frac{1}{R_1} - \frac{1}{R_2}\right)$$

$$= D\left\{\left(\frac{1}{R_1} + \frac{\nu}{R_2}\right) + 24(1-\nu)\frac{l^2}{h^2}\left(\frac{1}{R_1} - \frac{1}{R_2}\right)\right\}$$

(97a)

$$M_2 = M_{\sigma 2} + M_{\mu 2} = D\left(\frac{\nu}{R_1} + \frac{1}{R_2}\right) - 4Gl^2 h\left(\frac{1}{R_1} - \frac{1}{R_2}\right)$$

$$= D\left\{\left(\frac{\nu}{R_1} + \frac{1}{R_2}\right) - 24(1-\nu)\frac{l^2}{h^2}\left(\frac{1}{R_1} - \frac{1}{R_2}\right)\right\}$$

(97b)

These results clearly demonstrate that M-CST predicts inconsistent and unphysical results, which contradict with our mechanics intuition. Particularly, M-CST predicts:

- no size effect for pure spherical bending $R_1 = R_2$,
- significant size effect for pure anticlastic bending $R_2 = -R_1$,

On the other hand, when the pure bending deformation deviates from a pure equal curvature anticlastic deformation, where $\frac{1}{R_1} + \frac{1}{R_2}$ is non-zero, the couple-stress components appear in consistent couple stress theory (C-CST) and the flexural rigidity of the plate changes such that

$$M_1 = M_{\sigma 1} + M_{\mu 1} = D\left(\frac{1}{R_1} + \frac{\nu}{R_2}\right) + 4Gl^2 h\left(\frac{1}{R_1} + \frac{1}{R_2}\right)$$

$$= D\left\{\left(\frac{1}{R_1} + \frac{\nu}{R_2}\right) + 24(1-\nu)\frac{l^2}{h^2}\left(\frac{1}{R_1} + \frac{1}{R_2}\right)\right\}$$

(98a)

$$M_2 = M_{\sigma 2} + M_{\mu 2} = D\left(\frac{v}{R_1} + \frac{1}{R_2}\right) + 4Gl^2h\left(\frac{1}{R_1} + \frac{1}{R_2}\right)$$
$$= D\left\{\left(\frac{v}{R_1} + \frac{1}{R_2}\right) + 24(1-v)\frac{l^2}{h^2}\left(\frac{1}{R_1} + \frac{1}{R_2}\right)\right\} \quad (98b)$$

These results clearly demonstrate that the skew-symmetric consistent couple stress theory (C-CST) predicts consistent and physical results, which agree with experiments and our expectation. Most significantly, M-CST predicts:

- significant size effect for pure spherical bending $R_1 = R_2$,
- no size effect for pure anticlastic bending $R_2 = -R_1$.

For pure cylindrical bending $R_2 = \infty$, M-CST and C-CST predict the same total bending moments $M_1$, where

$$M_1 = (D + 4Gl^2h)\frac{1}{R_1} = D\left[1 + 24(1-v)\frac{l^2}{h^2}\right]\frac{1}{R_1} \quad (99)$$

However, these theories predict different bending moments $M_2$, where

$$M_2 = (vD - 4Gl^2h)\frac{1}{R_1} = D\left[v - 24(1-v)\frac{l^2}{h^2}\right]\frac{1}{R_1} \quad \text{in M-CST} \quad (100a)$$

$$M_2 = (vD + 4Gl^2h)\frac{1}{R_1} = D\left[v + 24(1-v)\frac{l^2}{h^2}\right]\frac{1}{R_1} \quad \text{in C-CST} \quad (100b)$$

When the thickness of the plate becomes $h = 2l\sqrt{\frac{6(1-v)}{v}}$, the total bending moment $M_2$ in M-CST disappears. By decreasing the thickness $h$ further, $M_2$ becomes negative. This non-physical result demonstrates once again an inconsistency in M-CST.

It should be mentioned that the pure bending of a beam of rectangular cross section with width $b$ and height $h$ in the $x_1$ direction can be approximated by the cylindrical bending of the flat plate, when $b \gg h$. This is because there are almost uniform normal force-stresses and couple-stresses acting on the transverse directions in the plate except near the lateral surfaces, where these out-of-plane stresses approach zero in thin boundary layers.

We notice that the formulations based on M-CST and C-CST predict the same in-plane deformations and stress distributions, where the bending moments $M_1$ are the same. However, the out-of-plane solution in M-CST is inconsistent and does not agree with expectations. As we illustrated, the bending moments $M_2$ in this theory can be positive, negative or zero depending on the length scale parameter $l$. This clearly shows that M-CST does not properly describe the pure bending of a beam, as well as that of a plate.

### 6.2. Consequences for the general plate theories

The modified couple stress theory (M-CST) has already been used to develop general size-dependent plate bending formulations in the framework of Kirchhoff-Love and Mindlin-Reissner plate theories. The basic assumption in a plate theory is that straight lines normal to the middle plane remain straight after deformation under arbitrary transverse loading. Kirchhoff-Love theory neglects the effect of transverse shear deformation by imposing the constraint that these straight lines remain normal to the middle deflection surface after deformation. On the other hand, Mindlin-Reissner theory includes the effect of transverse shear deformation without imposing any additional constraint. We notice that Kirchhoff-Love and Mindlin-Reissner plate theories represent pure bending precisely. This means that the solutions for pure bending deformation in these theories are exactly the same as analytical continuum mechanical solution. For example, for a linear isotropic elastic plate, the displacement components for pure bending in Kirchhoff-Love and Mindlin-Reissner theories are those given in Eq. (18).

Tsiatas (2009) and Şimşek et al. (2015) have developed Kirchhoff bending formulations for linear isotropic elastic plates based on M-CST. In these formulations, the general curvatures are the non-zero components of the torsion tensor $\chi_{ij}$

$$\chi_{11} = \frac{\partial^2 w}{\partial x_1 \partial x_2}, \qquad \chi_{22} = -\frac{\partial^2 w}{\partial x_1 \partial x_2}, \qquad \chi_{12} = \chi_{21} = \frac{1}{2}\left(\frac{\partial^2 w}{\partial x_2^2} - \frac{\partial^2 w}{\partial x_1^2}\right) \qquad (101\text{a-c})$$

where $w(x_1, x_2)$ is the transverse displacement of the middle plane of the plate. As demonstrated here, these formulations cannot possibly describe plate bending correctly. For pure bending of the plate into a pure spherical shell, the transverse displacement up to an arbitrary vertical rigid translation is approximated by the paraboloid

$$w = \frac{1}{2R_1}\left(x_1^2 + x_2^2\right) \qquad (102)$$

As expected, for this deformation, the components of the curvature tensor $\chi_{ij}$ (actually the torsion tensor) vanish, that is

$$\chi_{11} = 0, \qquad \chi_{22} = 0, \qquad \chi_{12} = \chi_{21} = 0 \qquad (103\text{a-c})$$

Therefore, there is no size effect and no couple-stresses for pure spherical bending, where

$$M_1 = M_2 = D(1+\nu)\frac{1}{R_1} \qquad (104)$$

Since M-CST cannot predict any size effect for this simple pure bending deformation, we must realize that these general plate bending formulations based on M-CST are inconsistent.

On the other hand, in the consistent couple stress theory (C-CST), the general curvatures are the non-zero components of the skew-symmetric mean curvature tensor $\kappa_{ij}$, where

$$\kappa_{12} = \frac{1}{2}\left(\frac{\partial^2 w}{\partial x_2^2} + \frac{\partial^2 w}{\partial x_1^2}\right) = \frac{1}{2}\nabla^2 w, \qquad \kappa_{21} = -\frac{1}{2}\left(\frac{\partial^2 w}{\partial x_2^2} + \frac{\partial^2 w}{\partial x_1^2}\right) = -\frac{1}{2}\nabla^2 w \qquad (105\text{a,b})$$

For the bending of the plate into the pure spherical shell, where the middle deflection surface is approximated by Eq. (102), these components are

$$\kappa_{12} = \frac{1}{R_1}, \qquad \kappa_{21} = -\frac{1}{R_1} \qquad (106\text{a,b})$$

As expected, C-CST predicts some size effect for pure spherical deformation, where

$$M_1 = M_2 = \left[D(1+v) + 8Gl^2h\right]\frac{1}{R_1}$$
$$= D\left[1+v+48(1-v)\frac{l^2}{h^2}\right]\frac{1}{R_1} \quad (107)$$

The inconsistency of M-CST can also be seen in Reddy et al. (2016) for axisymmetric bending, where the generally non-zero component of the torsion tensor in cylindrical polar coordinates centered at the center of plate is given as

$$\chi_{r\theta} = \chi_{\theta r} = \frac{1}{2}\left(-\frac{d^2w}{dr^2} + \frac{1}{r}\frac{dw}{dr}\right) \quad (108)$$

where $r = \sqrt{x_1^2 + x_2^2}$. We notice that for the pure spherical deformed surface of the middle plane, the transverse displacement $w(r)$ is the paraboloid

$$w(r) = \frac{r^2}{2R_1} \quad (109)$$

However, for this deformation, there is no corresponding curvature for M-CST, that is

$$\chi_{r\theta} = \chi_{\theta r} = 0 \quad (110)$$

and no couple-stresses.

The inconsistency of plate bending based on M-CST can also be observed in the framework of the Mindlin-Reissner plate theory. For example, the curvature components in Ma et al. (2011) vanish for pure bending of a plate into a spherical shell, where there is no shear deformation.

Although the development in this paper has been focused on isotropic elastic material, some of the results are still valid for specific anisotropic cases, such as orthotropic elastic plates. For example, the development of Tsiatas and Yiotis (2015) for orthotropic plates based on M-CST in the framework of Kirchhoff bending theory cannot describe the pure bending of the plate correctly. It turns out M-CST does not predict any size effect for pure bending of the orthotropic plate into a spherical shell. It should be emphasized that the inconsistency of M-CST is the direct result of the fact that the symmetric curvature tensor $\chi_{ij}$ in this theory is in reality a torsion tensor, which is not a proper measure of bending deformation.

Although we did not present the pure bending of plates in the original Mindlin-Tiersten-Koiter couple stress theory (MTK-CST), the solutions in that theory under infinitesimal deformation of linear elastic plates can be obtained by a linear combination of the solutions in M-CST and C-CST. However, we notice that all the inconsistencies with M-CST appear in these solutions. For example, MTK-CST also predicts size effect for the pure equal curvature anticlastic bending with $R_2 = -R_1$.

## 7. Conclusions

This paper shows that the pure bending of plates plays a very important role in examining the validity of the modified couple stress theory (M-CST) versus consistent couple stress theory (C-CST) from a practical point of view. In M-CST, the symmetric couple-stresses create torsion or anticlastic deformation with negative Gaussian curvature for surface elements of the continuum. As a result, this theory cannot describe pure bending properly. Particularly, M-CST predicts no couple-stresses and no size effect for the pure bending of a plate into a spherical shell. This result contradicts bending experiments and our intuitive expectation that there should be some size effects and an increase in flexural rigidity for this spherical bending deformation. This clearly shows that M-CST is inconsistent and the symmetric torsion tensor $\chi_{ij}$ is not a suitable measure of deformation in couple stress theory.

On the other hand in C-CST, the skew-symmetric couple-stresses creates ellipsoidal cap-like deformation with positive Gaussian curvature for surface elements of the continuum. As a result, the consistent couple stress theory (C-CST) describes pure bending of a plate properly. Particularly, C-CST predicts couple-stresses and some size effect for the pure spherical bending of the plate. This result completely agrees with bending experiments and our intuitive expectation. Therefore, this indicates that C-CST is consistent with physical reality and that the skew-symmetric mean curvature tensor $\kappa_{ij}$ is the suitable measure of bending deformation in couple stress theory.

We also notice that although M-CST and C-CST predict the same in-plane couple-stresses, deformation and flexural rigidity for cylindrical deformation of the plate, these theories predict different effects for out-of-plane deformation. This is because the couple-stresses in M-CST create anticlastic deformation with negative Gaussian curvature, whereas the couple-stresses in C-CST create ellipsoidal cap-like deformation with positive Gaussian curvature. This clearly demonstrates that M-CST cannot describe the pure bending of a beam properly.